 \documentclass[preprint]{aastex}

\usepackage{psfig}
\usepackage{times}
\usepackage{mathptm}

\newcommand{\kms}{\,\hbox{\hbox{km}\,\hbox{s}$^{\rm -1}$}}
\newcommand{\av}{\hbox{A$_{\rm V}$}}
\newcommand{\mv}{\hbox{m$_{\rm V}$}}
\newcommand{\mk}{\hbox{m$_{\rm K}$}}
\def\lk225{\hbox{Lk\,H$\alpha$\,225}}
\newcommand{\bd}{BD\,+40\arcdeg\,4124~}

\newcommand{\msun}{\,\hbox{M$_{\odot}$}}

\shorttitle{Adaptive Optics Integral Field Spectroscopy of \lk225}
\shortauthors{Davies et al.}

\begin{document}

\title{Adaptive Optics Integral Field Spectroscopy of the\\Young Stellar Objects in \lk225}

\author{R.I. Davies, M. Tecza, L.W. Looney, F. Eisenhauer, L.E. Tacconi-Garman, N. Thatte, \\ T. Ott, S. Rabien, }
\affil{Max-Planck-Institut f\"ur extraterrestrische Physik, Postfach 1312, 85741 Garching, Germany}

\author{S. Hippler and M. Kasper}
\affil{Max-Planck-Institut f\"ur Astronomie, K\"onigstuhl 17, 69117 Heidelberg, Germany}

\email{davies@mpe.mpg.de}


\begin{abstract}

Progress in understanding the embedded stars in \lk225 has been hampered by their variability, making it hard to compare data taken at different times, and by the limited resolution of the available data, which cannot probe the small scales between the two stars.
In an attempt to overcome these difficulties, we present new near-infrared data on this object taken using the ALFA adaptive optics system with the MPE 3D integral field spectrometer and the near-infrared camera Omega-Cass.
The stars themselves have K-band spectra which are dominated by warm dust emission, analagous to class I--II for low mass YSOs, suggesting that the stars are in a phase where they are still accreting matter.
On the other hand, the ridge of continuum emission between them is rather bluer, suggestive of extincted and/or scattered stellar light rather than direct dust emission.
The compactness of the CO emission seen toward each star argues for accretion disks (which can also account for much of the K-band veiling) rather than a neutral wind.
In contrast to other YSOs with CO emission, \lk225 has no detectable Br$\gamma$ emission.
Additionally there is no H$_2$ detected on the northern star, although 
we do confirm that the strongest H$_2$ emission is on the southern star, where we find it is excited primarily by thermal mechanisms.
A second knot of H$_2$ is observed to its northeast, with a velocity shift of -75\kms\ and a higher fraction of non-thermal emission.
This is discussed with reference to the H$_2$O maser, the molecular outflow, and [S{\sc ii}] emission observed between the stars.

\end{abstract}



\section*{}
\newpage

\section{Introduction}
\label{sec:int}

Herbig Ae/Be stars \citep{herbig} are commonly accepted as the 
intermediate mass (1.5--10\,\msun)
pre-main sequence counterparts to T Tauri stars.
\cite{hill92} separated Herbig Ae/Be stars into
three distinct categories based on their spectral
energy distributions, similar to T Tauri classifications.
However, the exact morphology that causes the infrared excess
is under debate, with most models assuming an accretion disk within a
large scale envelope \citep{natta93,hill92,james94}.
To further complicate matters, studies of these systems have shown that
binaries appear to be as common in Herbig Ae/Be systems as in T Tauri systems
\citep{leinert97,pirzkal97,testi98}. 
Thus, coeval star formation of Herbig Ae/Be stars,
or even Herbig Ae/Be stars with lower mass T Tauri stars, can provide
a wealth of information about intermediate star formation and
the role of these systems in clusters.

\bd (=V\,1685\,Cyg), one of the original Ae/Be stars of \cite{herbig}, 
is at the center of a partially embedded cluster of at least 33 stars with
infrared excesses \citep{strom72}.  The region, part of the giant star
forming region 2 Cyg at a distance of $\sim$~1kpc, has a significantly
younger stellar population than the nearby OB associations.  One of the
interesting aspects of this region is that both low and high mass stars are
forming simultaneously, possibly induced from an external shock into the
cloud core \citep{shev91,hil95}.  
Although optical images of the field around \bd are dominated
by the star itself, \lk225 (=V\,1318\,Cyg) is arguably one of the more
intriguing sources in the stellar cluster.
The \lk225 system is a possible embedded binary with apparent
separation of $\sim$5000 AU oriented north-south.
It has been proposed that the system consists
of both a young low mass star, \lk225~North, 
and an intermediate-mass very young Herbig Ae/Be star, \lk225~South,
based on luminosity arguments \citep{asp94}.
Many authors have suggested that the southern source is a center of star formation
activity based on a variety of tracers.
\cite{asp94} presented a 800\micron\ map which shows a strong peak on this source,
while \citet{pal95} found that both high density molecular gas traced by
CS\,$J=5$-4 and the total gas column density traced by C$^{18}$O\,$J=2$-1
are centered in this region.  More recently a 5\arcsec\ resolution
CS\,$J=2$-1 map by \citet{saf01} shows that the densest
gas is centred on the southern star of this pair.

In this paper we present adaptive optics observations, of J, H, and K
band imaging as well as integral field K band spectroscopy toward \lk225.
With the highest spatial resolution to date in the near-infrared for this
system we are able, for the first time, to make clear distinctions between
the emission from the two stars.  This allows us to highlight some intriguing
differences in the K-band emission from the two stars, as well
as a number of similarities.  Moreover, we present the results
of our unique ability to probe the inner binary region in a number 
of K-band emission lines.

\section{Observations \& Data Reduction}
\label{sec:obs}

The observations described here were taken using the ALFA adaptive optics (AO) system \citep{kas00} on the 3.5-m telescope at the German-Spanish Astronomical Centre on Calar Alto.
While ALFA has a laser guide star \citep{dav00}, here it was used in natural guide star mode. 
The wavefront reference was \bd (\mv=10.6, \mk=5.6), which lies 36\arcsec\ from \lk225 and is the only star in the group brighter than \mv=12.
While allowing a good correction on-axis, isoplanaticism at such a large angular distance is the main source of residual wavefront errors and severely limits the resolution achievable on our target.
The only real alternative is to use Lk\,H$\alpha$\,224 which is much closer.
However, while only 14\arcsec\ from the target it is too faint to correct on in the prevailing conditions;
its magnitude has been seen to vary in the range $12.5 \le V \le 17.2$ between 1985 and 1998 \citep{her99}. 
A discussion of the AO issues, with particular reference to the data obtained with Omega-Cass, is given in \citet{dav99}.

\begin{figure}
\centerline{\psfig{file=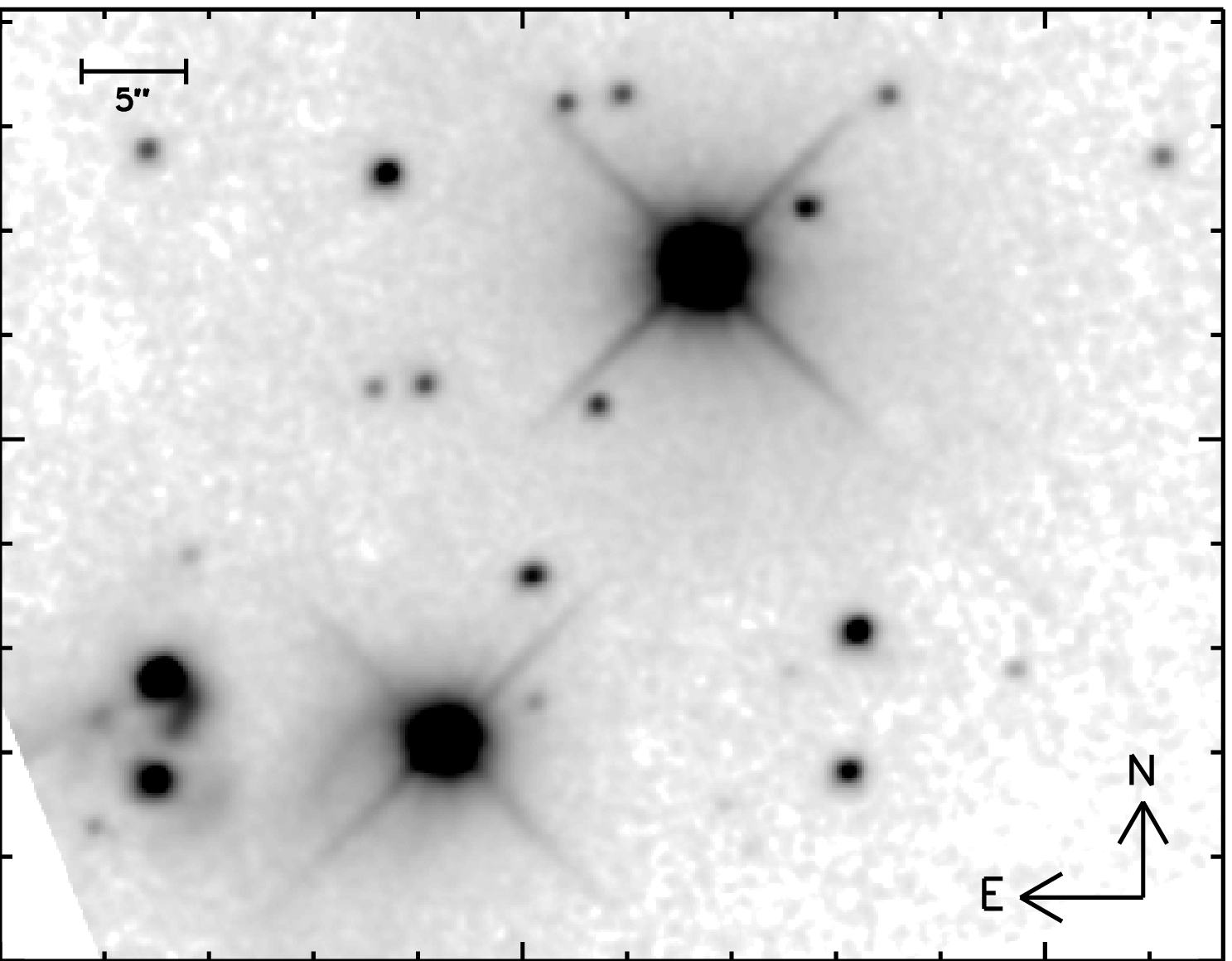,width=14cm}}
\caption{K-band image of the field around \bd, located top 
centre, plotted on a logarithmic scale \citep[from][]{dav99}.
The image has been smoothed with an adaptive filter so that bright sources 
are at full resolution while also allowing fainter sources to be seen.
The source 5\arcsec\ northwest of \bd is a ghost not an astrophysical object.
The other bright star is Lk\,H$\alpha$\,224, another Herbig Ae/Be star.
The double system \lk225 is the pair in the lower left of the
image, with an obvious arm of continuum emission curling between them.
These stars are 36\arcsec\ from \bd, which was used as
the wavefront reference as it is by far the brightest source in the V- and 
R-bands.
\label{fig:wide}}
\end{figure}

\begin{figure}
\centerline{\psfig{file=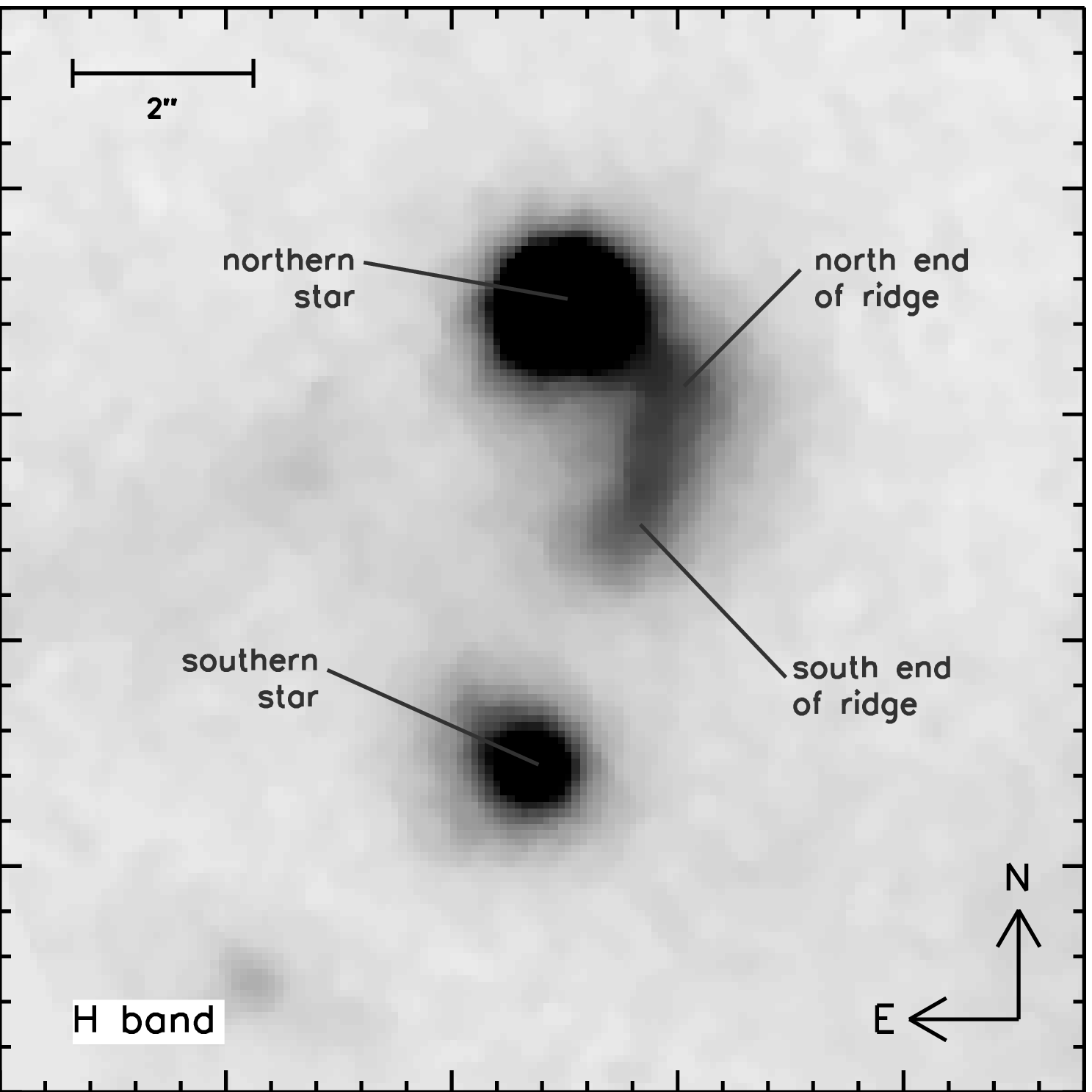,width=7cm}\hspace{5mm}\psfig{file=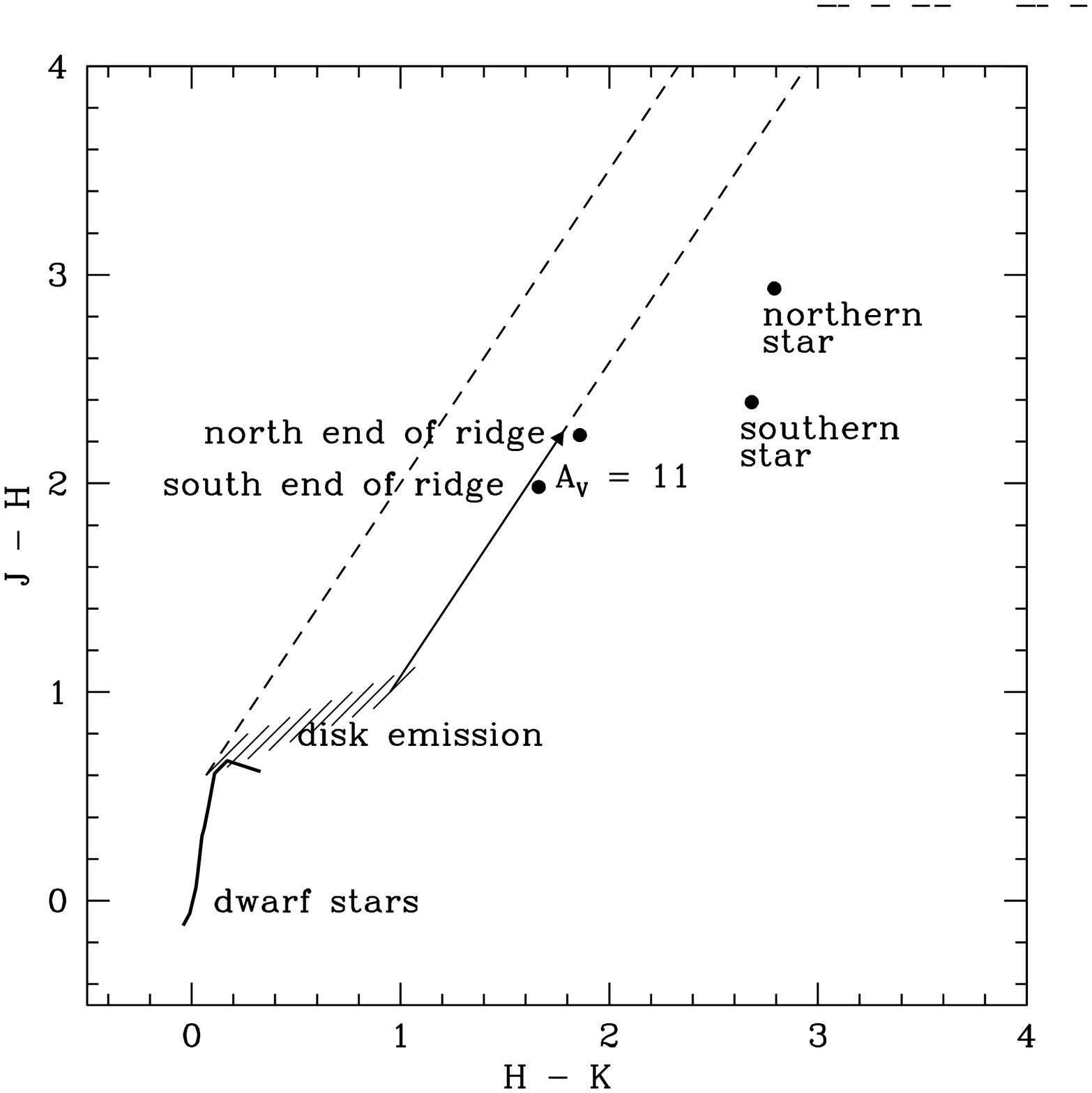,width=8.3cm,clip=}}
\caption{Left -- smoothed H-band image of the region around \lk225 (H-band is used as it has the best contrast for the ridge).
The ridge of continuum emission exhibits a sharp bend at its north end, which may suggest it does not in fact originate in the northern star.
Right -- colours of the 2 stars and 2 regions on the continuum ridge, as marked on the image, extracted in 0.8\arcsec\ apertures.
The curved line starting near (0,0) shows the colours of dwarf stars, and the hashed region indicates the effect of adding models for disk emission;
the reddening lines mark out the region where low and intermediate mass young stellar objects are expected to lie (adapted from \citealt{lad92}).
\label{fig:smal}}
\end{figure}

A K-band image of the field obtained with the near-infrared camera Omega-Cass in 1998, shown in Fig.~\ref{fig:wide}, has a FWHM resolution of 0.25\arcsec.
This and the corresponding J- and H-band images (Fig.~\ref{fig:smal}) were flux calibrated relative to star \#3 in \citet{hil95}.
This star is the brightest unsaturated star in the image ($K = 10.93$\,mag), but more importantly it is the least reddened with a $J-K$ colour of only 0.22\,mag, which tends to suggest it is the least susceptible to the variability associated with young stellar objects.

We have also obtained integral field spectroscopy using 3D \citep{wei96} during Oct\,1999 at a pixel scale of 0.25\arcsec, using AIM \citep{and98} as the interface between 3D and ALFA.
The pixel scale, ambient seeing conditions ($\sim$1\arcsec), and distance to the reference star limited the final resolution to $\sim$0.6\arcsec.
3D is an integral field spectrograph that simultaneously obtains spectra
for each of 256 spatial pixels covering a square field of view.  
The spectral wavelength range spans from
$1.912\micron$ to $2.382\micron$ at a spectral resolving power
($\mathrm{R}\equiv\lambda/\Delta\lambda$) of 1000.  Two pixel scales
can be selected with AIM, $0\farcs07$ and $0\farcs25$. 
The latter was used for these observations, giving a field of
$4\arcsec\times 4\arcsec$.  To observe both \lk225\,N and~S, two pointings 
were necessary.  The total on-source integration time for each was 
$600\,\mathrm{s}$ with individual frame
integration times of $100\,\mathrm{s}$.  The same amount of time was spent
off-source for sky background subtraction.
The data were reduced using the set of 3D data analysis routines written
for the GIPSY \citep{VHU1992} package.  This included
wavelength calibration, spectral and spatial flat fielding, dead and hot
pixel correction, and division by the reference stellar spectrum.  Data
cubes from the individual exposures were recentered and added using the
centroid of the broadband emission.  A mosaic of the two pointings was
created using the relative coordinates of \lk225\,N and~S
given in \citet{hil95}.  Emission line and absorption
line maps were extracted by subtracting a spline fit to the entire line-free 
continuum in each spatial pixel.

\section{Morphology \& YSO class}
\label{sec:dot}

The system \lk225 comprises 
the pair of stars in the south-east corner of Fig.~\ref{fig:wide}, which are separated by 5\arcsec, about 5000\,AU at the 
distance of the cluster \citep[980\,pc,][]{hil95}.
While this is rather large, nearly 5\% of known YSOs do have
such wide separations \citep{mat94}, although it is not certain whether they are gravitationally bound or simply have formed close to each other from the same cloud core.

\begin{figure}
\centerline{\psfig{file=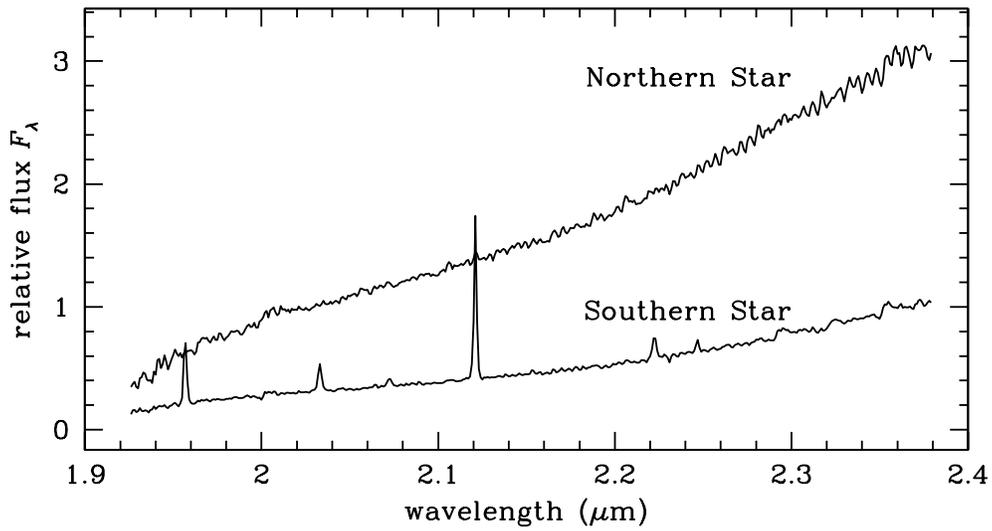,width=14cm}}
\caption{K-band spectra of the 2 stars in \lk225, extracted in 0.88\arcsec\ boxes.
The southern star exhibits strong H$_2$ emission, while the northern
star has none.
The continuum in both cases is well matched by a 450--500\,K
blackbody and a 20--40\% contribution from an extincted A-type star.
\label{fig:spec}}
\end{figure}

The strong variability of this object, brightness changes of more than 2\,mag in V and V-R colour changes of more than 2\,mag within 1 year \citep{ibr88}, means that it is not straight forward to compare observations at different times.
Based on an R-band image, \citet{asp94} identified 3 components to \lk225 in a north-south line and a total separation of 5\arcsec.
The bright central component was also seen in the R-band image of \citet{mag97}, distinct from both \lk225\,N and~S.
It is difficult to attribute this to a third star because the infrared appearance is quite different, with a long nebulous tail extending from the northern star.
Fig.~\ref{fig:smal} shows that the tail is rather straight with a sharp kink at the north end, which may suggest it does not originate in the northern star but only appears to because it lies so close.
The magnitudes of several points around \lk225 (Table~\ref{tab:mags}) and the colours marked in the JH-HK diagram indicate that not only is the whole ridge bluer than the two stars, but that the south end is bluer than the north end.
If this is due to a decrease in extinction along the ridge of \av=2\,mag, it may explain why R-band images give the impression of a third discrete source.
The colours along the ridge are easily achieved by reddening stellar light with a disk emission component by \av$\sim$11.
This is in contrast to the two stars which have a much redder $H-K$ colour.
Anomalous extinction laws have been observed towards a number of Herbig Ae/Be stars \citep[][and references therein]{stee89,stee91} which can modify the direction of the reddening in the colour diagram, calculations by these authors have shown that this change is towards bluer H-K colours.
Thus the position of the stars in the diagram appears to demand an additional significant (60--80\%) dust contribution, which must arise in an envelope rather than a disk.
Their position in the diagram suggests they are Class~I young stellar objects (to which the Group~I Herbig Ae/Be stars of \citet{hill92} appear to be analogous), or intermediate between Class~I and~II.
Apart from Class~0 sources (which do not emit at wavelengths shorter than 10\micron\ due to heavy extinction) these are the least evolved sources, with accretion disks, and surrounded by envelopes of gas and dust.
The spectra presented in Fig.~\ref{fig:spec} tend to support this interpretation, having shapes suggestive of blackbody emission at $\sim600$\,K, most similar to the Class~I objects in the spectral atlas of \citet{gre96}.
Such objects are associated with very high K-band veilings -- defined as the ratio of excess to photospheric emission in the K-band, 
$r_{\rm k} = F_{\rm Kex}/F_{\rm K*}$. 
The excess emission is believed to come from hot dust or an accretion
disk around the star and does not necessarily redden the star itself.
Flares in such an accretion disk could explain the K-band variability of the two stars reported by various authors:
the northern source was the brighter of the two in Sep~93 \citep{hil95}, Nov~93 \citep{asp94}, and both Aug~98 and Oct~99 (this paper); while the southern star was brighter during Oct~91 and May~93 \citep{asp94}.

One possibility to explain the emission between the two stars is that the cloud core from which \lk225 formed condensed into a long filament. 
The continuum we see is tracing the remnants of this filament, arising either through direct emission from a string of young low-mass stars, or more likely since we do not see discrete sources, through light scattered from such stars.
Polarisation studies would help to make progress on understanding this unusual morphology.

\section{H$_2$ line emission}
\label{sec:h2}

\begin{figure}
\vspace{1cm}
\centerline{\psfig{file=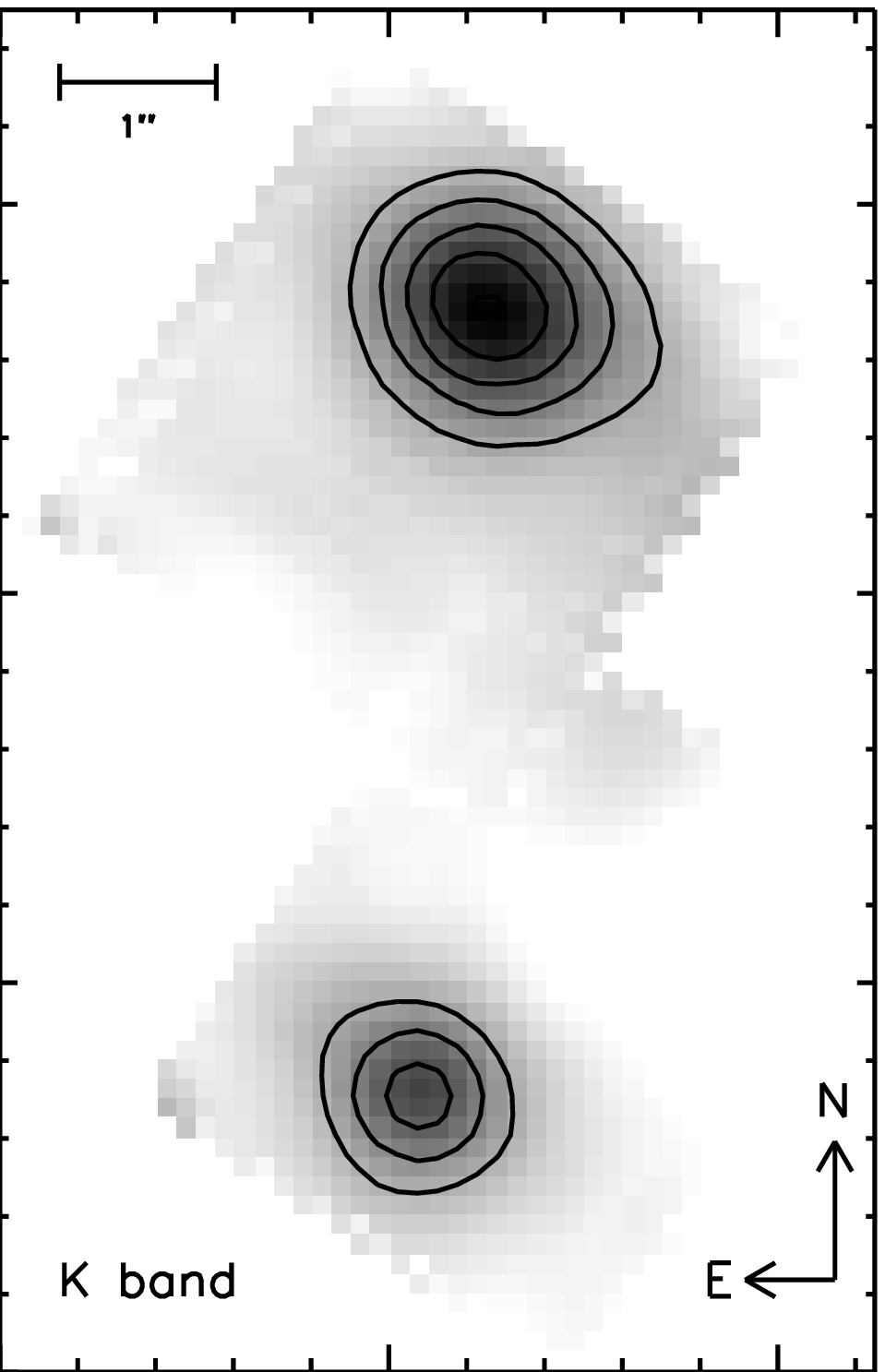,width=5cm}\hspace{5mm}\psfig{file=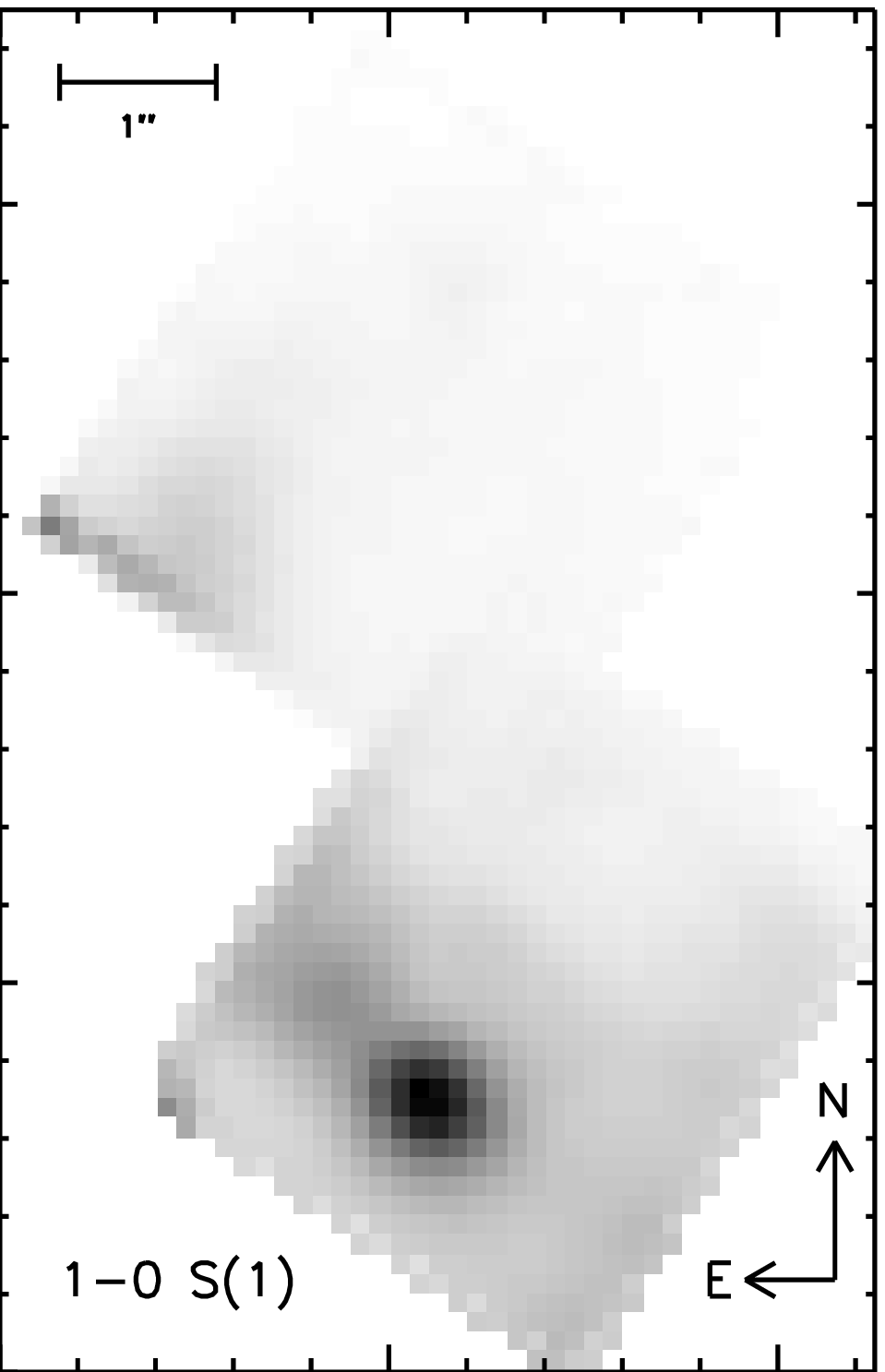,width=5cm}\hspace{5mm}\psfig{file=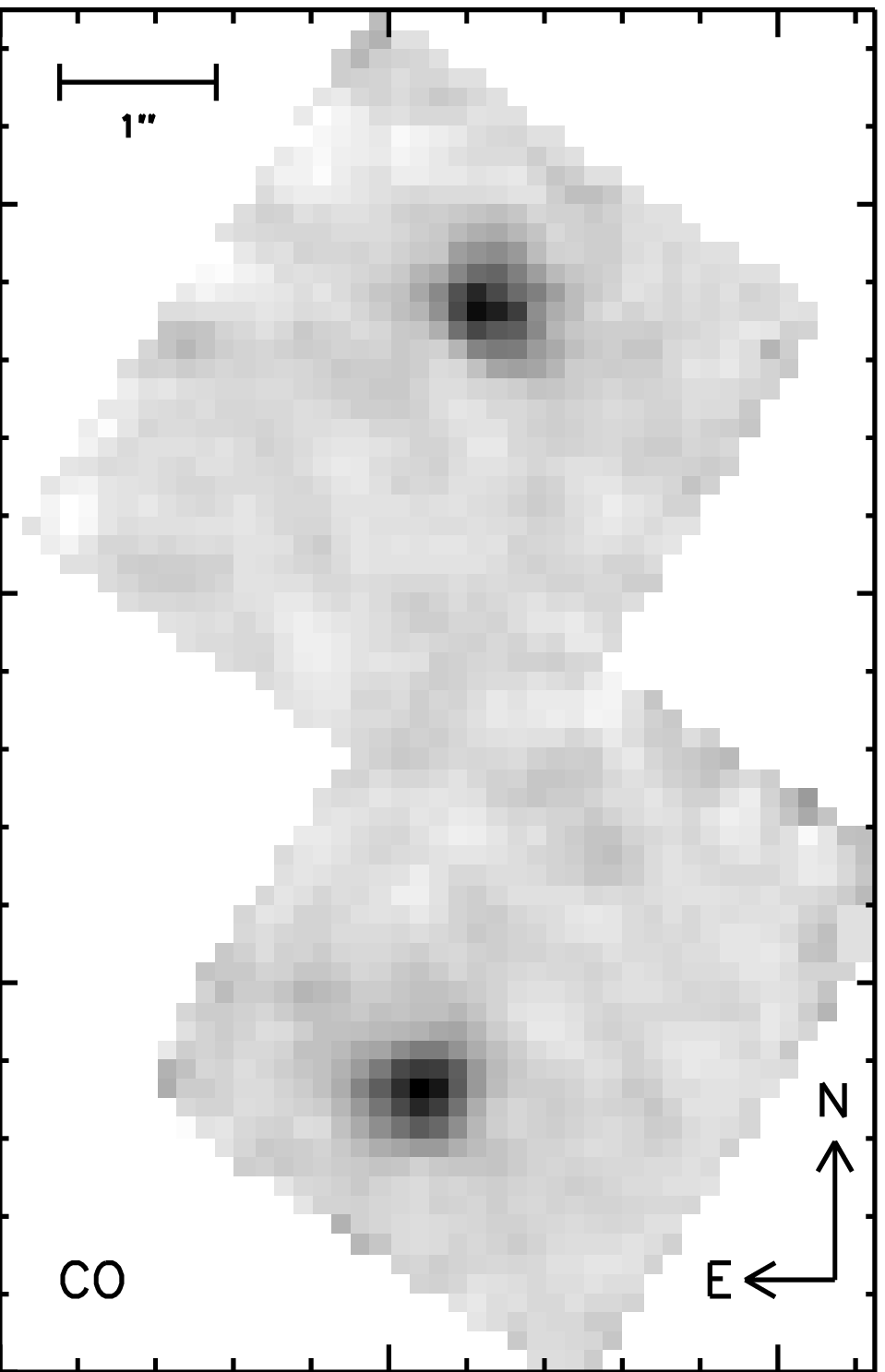,width=5cm}}
\caption{Images of \lk225 observed with 3D and ALFA.
Due to non-ideal observing conditions and the
distance to the reference, the resolution achieved was $\sim$0.6\arcsec.
Left: K-band (sum of all spectral channels) continuum showing faintly the emission curving to the west between the stars.
Centre: H$_2$ 1-0\,S(1) line map -- there is strong emission both on
the southern star and to its east, as well as some diffuse emission.
Right: CO 2-0 bandhead 2.3\micron\ line emission map -- the fluxes are similar, but the CO equivalent width for the northern star is much less.
\label{fig:line}}
\end{figure}

The advantage of integral field spectroscopy is the ability to obtain line maps and spectra simultaneously.
Fig.~\ref{fig:line} (centre) is a map of the H$_2$ 1-0\,S(1)
emission confirming what the spectra show, that the strongest emission is located directly on the southern star (we find no difference in their centroid positions), with a ratio S(1)/Br$\gamma$$>$20.
Additionally, there is an outflow/jet or extra source 1\arcsec\ to its north-east.
For reference, the left image (both greyscale and contours) is the integrated K-band continuum to indicate the positions of the 2 stars on the field.

\citet{asp94} observed weak emission from the northern star, while we have detected none.
It is possible that a combination of seeing and large pixel size (3\arcsec) meant that some of the flux from the southern star bled over to the pixels on which the northern star is located.
With a finer pixel scale and high resolution our data do not suffer from this problem.

\begin{figure}
\centerline{\psfig{file=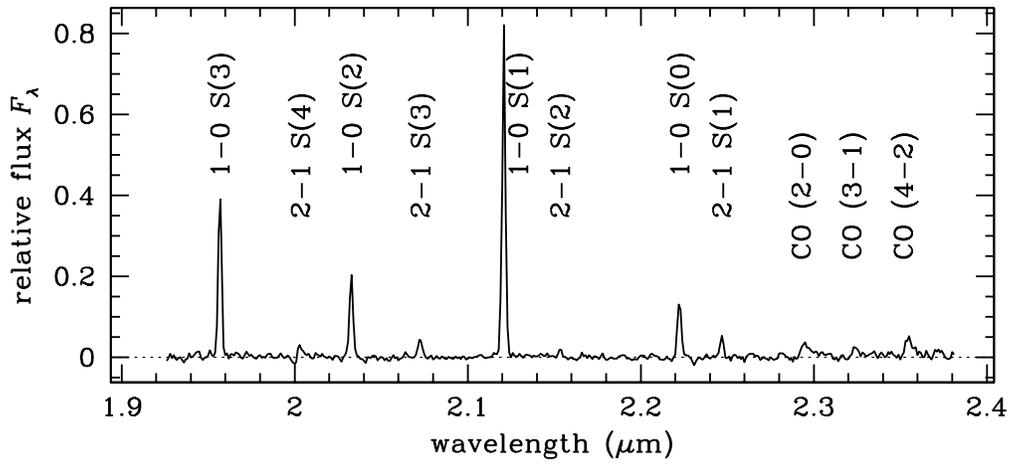,width=14cm}}
\caption{Line spectrum of \lk225\,S after subtracting the continuum, with the features marked.
\label{fig:h2sp}}
\end{figure}

\begin{figure}
\centerline{\psfig{file=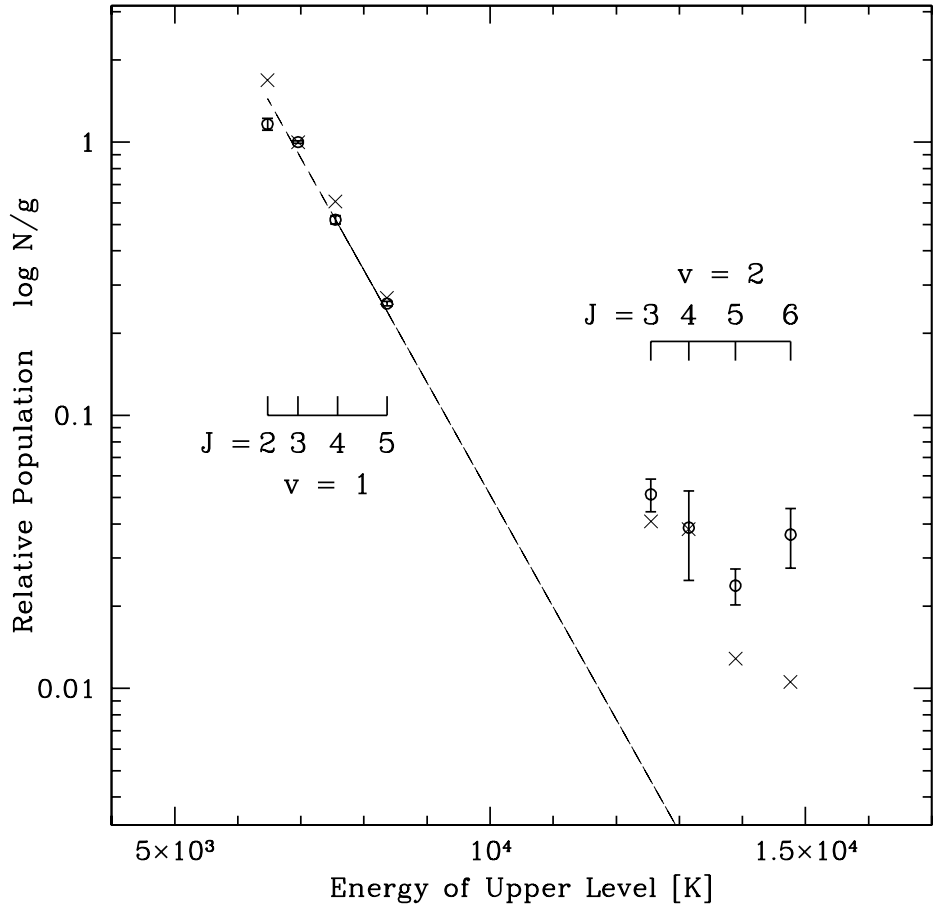,width=13cm}}
\caption{Population diagram for the hot H$-2$ molecules in \lk225\,S, assuming a LTE ortho/para ratio of 3 and no extinction.
While the assumption about ortho/para ratio is not necessarily correct, the diagram still shows clearly that the $\nu=1$-0 transitions are dominated by thermal processes, while the $\nu=2$-1 are predominantly non-thermal.
The crosses are a best fit model to the spectrum, comprising a combination of thermal emission and UV fluorescence (assuming a density below 10$^4$\,cm$^{-3}$).
The dashed line is the purely thermal part, which has a characteristic temperature of 1000\,K.
\label{fig:h2po}}
\end{figure}

A continuum-subtracted spectrum of the south star is presented in
Fig.~\ref{fig:h2sp}, in which 8 H$_2$ lines are detectable at 
$\ga$3$\sigma$, 4 from each of the $\nu$=1-0 and $\nu$=2-1 series.
The relative line fluxes are summarised in Table~\ref{tab:h2flux},
and in Fig.~\ref{fig:h2po} we have derived a relative population diagram for hot H$_2$ molecules from the lines assuming the LTE value of 3 for the ortho/para ratio.
While this assumption is not necessarily strictly correct, the diagram is a very useful presentation that gives an indication of whether the emission is thermal.
We have further assumed that there is negligible differential extinction between 1.95--2.25\micron, in contrast to the \av=25$\pm$9\,mag estimated by \citet{asp94} from the ratio of the lines 1-0\,Q(3) to 1-0\,S(1).
We find that for \av$\gtrsim$9 implies the higher rotational levels have larger populations (even for the $\nu=1$ vibrational level which is derived from lines with high signal-to-noise).
We suspect that \citet{asp94} may have underestimated the continuum level longward of 2.35\micron, as this is hard to determine at low spectral resolution due to the many H$_2$ and CO emission lines:
our spectra in Fig.~\ref{fig:spec} show that the continuum begins to rise rapidly at the long end of the K-band, the effect of which is to reduce the flux estimated in the Q-branch lines and thus also the derived extinction.

From an analysis of H$_2$ lines in an ISO spectrum, \citet{anc00} concluded that there were two components: a C-shock (thermal component) with only mild (\av$\sim$15) extinction, and a J-shock (non-thermal component) that is heavily (\av$\sim$50) extincted.
A reconciliation of these high extinctions with the low ones implied by our data may be possible if there is a significant extended H$_2$ component, perhaps which we are beginning to see in Fig~\ref{fig:line}.

Without any correction for extinction the $\nu=1$ levels are well approximated by thermal excitation at a temperature of T$_{ex} = 1000$--1100\,K.
Correcting for an assumed \av=5\,mags of extinction raises the temperature derived from the $\nu=1$ levels to nearly 2700\,K.
In order to make the most of even the weaker lines which have fluxes that are hard to measure we have taken the following approach.
We have calculated the theoretical spectrum from 1.93--2.28\micron\ (to avoid the CO emission) using a combination of thermal (ortho/para ratio 3) and non-thermal components.
For the latter, model 14 from \citet{bla87} (with ortho/para ratio $\sim$1.7 and calculated for a density of $3\times10^3$\,cm$^{-3}$ and a incident UV radiation field of $3\times10^3$ times the background) was used.
Since the relative line intensities are fairly insensitive to the density and strength of the radiation field, it provides a typical spectrum over the range of conditions that were considered;
thus while we can estimate the fraction of H$_2$ emission due to fluorescence, we cannot constrain the conditions for its excitation.
Additionally, in our fitting procedure we have left the FWHM and velocity shift of the lines as free parameters.
The results of the $\chi^2$ minimisation are given in Table~\ref{tab:h2}.
In summary, we find that the H$_2$ lines at \lk225\,S are unresolved and have zero velocity shift; thermal excitation at 1070\,K contributes 91\% of the 1-0\,S(1), and the remainder has a non-thermal origin.
Two temperature models are ruled out since a higher temperature component that is prevalent enough to account for the $\nu=2$-1 lines would significantly modify the observed fluxes in some of the $\nu=1$-0 lines.

The calculation here assumes the low density regime, but on the other hand dense gas is a prerequisite for H$_2$O masers, one of which is seen in this region \citep{pal95}.
The population diagram is indeed reminiscent of the high density photo-dissociation region (PDR) models of \citet{ste89}, where H$_2$ excitation occurs above the critical density at which collisional effects dominate the spectrum.
These authors found that for UV intensities more than 100 times the background and densities above $10^5$\,cm$^{-3}$, the low vibrational levels become thermalised with a characteristic temperature of 1000\,K.
However, their models showed that this extends to the $\nu$=2 level, exhibited as thermal ratios for the $\nu=2$-1 as well as the $\nu=1$-0 lines. 
This appears not to be the case for our data (Fig.~\ref{fig:h2po}), arguing against the scenario.

We have extracted a spectrum of the H$_2$ emitting region to the northeast of \lk225\,S, in a 0.5\arcsec$\times$0.5\arcsec\ box.
The same approach indicates that H$_2$ lines in this region are resolved, with a quadrature corrected FWHM of 220\kms, and are shifted by -75\kms.
Only 81\% of the 1-0\,S(1) has a thermal origin, with a characteristic temperature of 980\,K.
That 19\% of the 1-0\,S(1) lines has a non-thermal origin implies that more than half of all the H$_2$ cooling that can be deduced from these observations is through fluorescent emission.
We cannot rule out the possibility that the broad line is due to two narrow components with different velocity shifts;
but in any case there must be a significant component at -75\kms.
This result is discussed in Section~\ref{sec:out} with reference to the [S{\sc ii}] line emission and the molecular outflow.

\section{CO emission}
\label{sec:co}

The 2.3\micron\ CO~2-0 bandhead line map in Fig.~\ref{fig:line}~(right) clearly indicates that there is a compact spot of CO emission centered on each of the two stars and none elsewhere.
In an analysis of the CO line emission seen with ISO, \citet{anc00} used four rotational $\nu=0$-0 transitions between 150--190\micron, which have very low excitation temperatures and so trace cool gas heated thermally rather than fluorescently; hence they derived a thermal excitation temperature of only 300\,K.
If both the rotational and vibrational transitions of CO are excited collisionally (as argued by \citealt{car89}) then the CO~2-0 bandheads at 2.3\micron\ indicate an excitation temperature of more than 3000\,K and a density greater than $10^{10}$\,cm$^{-3}$.

\citet{car89}, using their own and data from \citet{geb87}, demonstrated that all the CO emitting sources then known also exhibited Br$\gamma$ emission, and that there was a correlation between them, the Br$\gamma$ luminosity being a factor of 2 or so greater than the CO~2-0 luminosity.
Understanding this relationship has been difficult since the converse is not true, and only 25\% of Br$\gamma$ emitters also exhibit CO emission.
One unusual aspect of \lk225\,S is that it breaks this correlation by exhibiting a ratio CO\,2-0/Br$\gamma > 5$ (we have not presented a Br$\gamma$ map since there is no detectable emission in 3D's entire field of view).
Although \citet{tho85} proposed a model in which both CO and Br$\gamma$ emission originate in the same gas, \citet{car89} argue it is more likely that there are two components, hot (partially) ionised gas and cool ($\lesssim4000$\,K) neutral gas.
Our observations can be explained if there is no hot component (or it is highly extincted), or if the temperature of Thompson's model is lowered.
The concept of different lines arising in different gas components can be extended to include the H$_2$ emission in the southern star.
The conditions needed to observe this line ($1000\lesssim{\rm T}\lesssim3000$\,K, density $\lesssim10^{\rm 6-7}$\,cm$^{-3}$) are rather different to those for the CO, and so it must also arise in a different volume of gas.

We cannot use the CO line profile from the observations presented here to distinguish between accretion disk or neutral wind models, but instead the CO morphology can help.
The molecular outflow seen by \citet{pal95}, for which they estimated a mass loss rate of $3\times10^{-3}$\msun\,yr$^{-1}$, would tend to support the wind model.
To verify or veto this it is crucially important to know where the outflow originates. 
CO emission is seen only on the stars, inconsistent with an outflow that begins between them;
indeed two distinct sources of CO emission is inconsistent with any model in which the CO arises in a single outflow.
Additionally, the mass loss rate is also several orders of magnitude larger than models of winds predict from CO emission \citep{car89}.
The alternative accretion disk model would naturally explain the compactness of the CO emission; 
and it also appears that it could simultaneously account for both the veiling and CO emission, at least in the southern source. 
In the northern source the CO flux is similar and could indicate
similar conditions (density, temperature, accretion rate) in an
accretion disk, but an additional warm dust envelope probably
contributes to the higher veiling.

\section{Shocks, the Maser, \& the Molecular Outflow}
\label{sec:out}

From observations of the $^{12}$CO\,2-1 line at 1.3\,mm, \citet{pal95} discovered a molecular outflow originating in \lk225.
They identified the origin of this with the H$_2$O maser in this object which they confirmed to have a velocity shift of -80\kms, and which appears to lie on the southern star.
The coordinates given by \citet{pal95} and \citet{hil95} (B1950 and J2000 respectively) do not agree to better than a few arcsec, so we have used the 2MASS astrometry (accurate to $<$0.2\arcsec) combined with our K-band image to determine accurate an position for \lk225\,S ($\alpha=20\,20\,30.59$, $\delta=41\,21\,26.0$, J2000).
The maser lies $0.4\pm0.2$\arcsec\ north and $0.1\pm0.2$\arcsec\ east of this star.
If the transverse velocity of the maser is similar to its radial velocity, then it will not have moved by more than $\sim$0.1\arcsec\ between the radio and our near-infrared observations.

A number of different observations add to the complexity of the source.
We have shown that there is a second weaker region of H$_2$ emission to the northeast of \lk225\,S.
It is also only 0.6\arcsec\ from the maser, and has a broader line width that is shifted by -75\kms, the same velocity as the maser.
These are all suggestive of a link between the maser and H$_2$ source.
An important argument against the much brighter H$_2$ emission on the star itself being associated with the maser is that it shows no such velocity shift.
A second line of reasoning is based on the maser model proposed by \citet{eli89} in which H$_2$O maser could occur in the $\sim$400\,K temperature plateau behind a J-shock.
\citet{eli95} further showed that C-shocks can also give rise to such masers, increasing the phase space to include preshock densities of 10$^6$--10$^8$\,cm$^{-3}$ and velocities 20--200\kms.
The spectrum of the H$_2$ knot to the east of \lk225\,S shows that more than half of the total H$_2$ emission arises from molecules that have been non-thermally excited (a greater fraction than in the H$_2$ emission coincident with the star). 
One way to achieve this is through H$_2$ dissociation in a J-shock, if the shock velocity is more than 30--40\kms\ -- although not faster than this since significant Br$\gamma$ emission must be avoided.
It is the heat from re-formation of the molecules behind the shock front that produces the warm plateau ideal for maser formation \citep{eli89}.
Based on these arguments, we suggest that this H$_2$ knot and the H$_2$O maser may be excited in the same shock system.

The other emission which could be associated with the maser is the strong [S{\sc ii}], which is also shifted by -80\kms\ with respect to \lk225-N, and was observed between the two stars by \citet{mag97}.
In this region they found almost no detectable H$\alpha$ or [N{\sc ii}] emission, 
putting tight constraints on the excitation temperature since the ionisation potential of hydrogen and nitrogen is 13.6\,eV and 14.5\,eV respectively, while for sulphur it is 10.4\,eV.
In their later observations (1995 rather than 1992) they also saw [S{\sc ii}] had appeared at the location of the northern star.
Such rapid variability in the emission lines effectively rules out shock-excitation since models show that the flow time through the hot shock structure is on the order of 100\,yrs.
We would argue that if this variability of the [S{\sc ii}] is genuine then it is more likely to be in gas photo-ionised by one or both of the young stars (as discussed earlier, at least one may be late-B or early-A type).
The -80\kms\ velocity shift could be reproduced if the ionised gas was swept up in the outflow (while the average outflow velocity is 8\kms\ \citep{pal95}, it could be much higher nearer the stars).
The variations might then be tied to changes of optical depth between the star and the gas, or be associated with flaring up of the accretion disk which would act as an intermittent source of higher energy photons necessary to produce [S{\sc ii}].

\section{Conclusions}
\label{sec:con}

We have presented adaptive optics imaging and integral field spectroscopy of the deeply embedded and highly variable sources associated with \lk225 in the star forming complex around \bd.
The 2 bright stars have JHK colours and K-band spectra typical of Class~I-II YSOs, with an additional component of emission coming from a more extended envelope rather than just a circumstellar disk.
The extinction to the stars is uncertain, but seems to be \av$\sim$11\,mag.
The ridge of continuum emission between the stars has bluer colours than the stars themselves, inconsistent with direct dust emission but appears instead to be stellar light that is extincted and perhaps also scattered.

There are 2 sources of H$_2$, one co-incident with the southern star, and one 1\arcsec\ to its north-east.
For their line ratios to be physically meaningful neither can be reddened by more than \av$\sim$9\,mags.
In both cases most of the 1-0\,S(1) emission has a thermal origin with a characteristic temperature of 1000\,K, but the latter source does have a greater contribution from non-thermal emission.
Its lines are broader and shifted by -75\kms, suggesting it is this knot which is associated most closely with the water maser (astrometric measurements place the maser almost equidistant from the star and this region).
The variability reported by some authors in the [S{\sc ii}] lines, which occur between the star and near the northern star, would seem to rule out shock excitation for this line.
We suggest that it is a result of gas (which itself is caught up in the outflow) photo-ionised by one of the young stars, and that flares in the accretion disk or changes in optical depth to the star cause the variations in line flux.

Both stars show CO lines in emission, but in contrast to other CO emitting YSOs, there is no detectable Br$\gamma$ emission.
The molecular outflow seen in the radio CO emission provides ample evidence for a neutral wind which could also give rise to 2.3\micron\ CO emission.
However, two distinct 2.3\micron\ CO sources is not consistent with a model of a single outflow;
determining the exact origin of the outflow with respect to the K-band continuum emission would be a big step towards understanding the emission.
On the other hand, 
an accretion disk model would naturally explain the compactness of the CO emission regions on each star, and it could account for at least some of the veiling seen.


\acknowledgments

The authors are grateful to both the ALFA and 3D teams for their invaluable help, and the Calar Alto staff for their hosptality.
We extend our thanks to Mansur Ibrahimov for providing a copy, and translation, of his paper.
RD acknowledges the support of the TMR network `Laser Guide Stars for 8-m Class Telescopes'.
This publication makes use of data products from 2MASS, a joint project of the University of Massachusetts and IPAC/CalTech, funded by NASA and the NSF.


\bibliographystyle{natbib}

\clearpage

\begin{deluxetable}{rccc}
\tablecaption{Magnitudes of \lk225\label{tab:mags}}
\tablewidth{0pt}
\tablehead{
 \colhead{position} &
 \multicolumn{3}{c}{magnitude} \\
 \colhead{} &
 \colhead{K} &
 \colhead{H} &
 \colhead{J}
}
\startdata
North star		& 10.0 & 12.7 & 15.7 \\
South star		& 11.5 & 14.2 & 16.6 \\
North end of ridge	& 13.1 & 14.9 & 17.2 \\
South end of ridge	& 13.5 & 15.2 & 17.2 \\
\enddata
\end{deluxetable}

\begin{deluxetable}{cccccc}
\tablecaption{Relative fluxes for observed H$_2$ lines\label{tab:h2flux}}
\tablewidth{0pt}
\tablehead{
 \colhead{Line} &
 \colhead{wavelength} &
 \colhead{$E_u$/$k$\,\tablenotemark{a}} &
 \colhead{$A_{ul}$\,\tablenotemark{a}} &
 \multicolumn{2}{c}{Relative Flux} \\
 \colhead{} &
 \colhead{(\micron)} &
 \colhead{(K)} &
 \colhead{($10^{-7}$\,s$^{-1}$)} &
 \colhead{Southern star\,\tablenotemark{b}} &
 \colhead{Eastern emission\,\tablenotemark{c}}
}
\startdata
1-0\,S(3) & 1.9576 & \phn8365 & 4.21 & \phn5.30 & \phn5.02 \\
1-0\,S(2) & 2.0338 & \phn7584 & 3.98 & \phn2.67 & \phn3.53 \\
1-0\,S(1) & 2.1218 & \phn6956 & 3.47 &    10.00 &    10.00 \\
1-0\,S(0) & 2.2235 & \phn6471 & 2.53 & \phn1.93 & \phn2.98 \\
2-1\,S(1) & 2.2477 &    12550 & 4.98 & \phn0.70 & \phn1.35 \\ 
2-1\,S(2) & 2.1542 &    13150 & 5.60 & \phn0.26 & \phn0.30 \\
2-1\,S(3) & 2.0735 &    13890 & 5.77 & \phn0.64 & \phn0.96 \\
2-1\,S(4) & 2.0041 &    14764 & 5.57 & \phn0.38 & \phn0.75 \\ 
\enddata
\tablenotetext{a}{taken from the UKIRT Astro-Utilities page}
\tablenotetext{b}{ratios normalised to 1-0\,S(1), with 1\,$\sigma$ errors $\sim$0.17}
\tablenotetext{c}{ratios normalised to 1-0\,S(1), with 1\,$\sigma$ errors $\sim$0.12}
\end{deluxetable}

\begin{deluxetable}{rrr}
\tablecaption{Derived parameters for H$_2$ emission\label{tab:h2}}
\tablewidth{0pt}
\tablehead{
 \colhead{Quantity} &
 \colhead{Southern star} &
 \colhead{Eastern emission}
}
\startdata
temperature of thermal emission		& 1070\,K	& 980\,K	\\
non-thermal fraction of 1-0\,S(1)	&  9\%	& 19\%	\\
non-thermal fraction of total H$_2$	& 38\%	& 59\%	\\
FWHM of lines			  	& 305\kms	& 375\kms	\\
velocity shift			  	& -5\kms	& -75\kms	\\
\enddata
\end{deluxetable}

\clearpage

\end{document}